\documentclass[a4paper]{llncs}
\usepackage{url}
\usepackage{times}
\usepackage{color}
\usepackage{graphicx}
\usepackage{amsmath}
\usepackage{float}
\usepackage{url}
\usepackage{hyperref}
\usepackage[linesnumbered,lined,algoruled,algochapter,dotocloa]{algorithm2e}
\usepackage{caption}
\usepackage{subcaption}
\usepackage{listings}
\usepackage{amsmath}
\usepackage{amssymb}
\usepackage{tikz}
\usetikzlibrary{arrows,automata,positioning,trees,shapes,
decorations.pathreplacing,trees}

\newcommand{\comment}[1]{}

\begin{document}

\title{Beyond $k$-induction: Learning from Counterexamples to Bidirectionally
Explore the State Space}

\author{Mikhail R. Gadelha$^1$ \and
        Felipe R. Monteiro$^2$ \and
        Enrico Steffinlongo$^1$ \and\\
        Lucas C. Cordeiro$^{3}$ \and
        Denis Nicole$^1$}
\authorrunning{M. R. Gadelha, F. R. Monteiro, L. C. Cordeiro, and D. Nicole}
\institute{
  $^1$University of Southampton, United Kingdom\\
  $^2$Federal University of Amazonas, Brazil\\
  $^3$University of Manchester, United Kingdom \\
  \url{esbmc@googlegroups.com }
}
\maketitle

\begin{abstract}
We describe and evaluate a novel $k$-induction proof rule called bidirectional
$k$-induction (\textit{bkind}), which substantially improves the $k$-induction
bug-finding capabilities. Particularly, \textit{bkind} exploits the
counterexamples generated by the over-approximation step to derive new
properties and feed them back to the bounded model checking procedure. We also
combine an interval invariant generator and \textit{bkind} to significantly
improve the number of correct verification results. Experimental results show
that \textit{bkind} can considerably reduce the verification time compared to
the na\"ive $k$-induction proof rule, since it only requires half the number of
steps to find a given safety property violation in an unsafe program. The
\textit{bkind} algorithm outperforms 2LS, another state-of-the-art $k$-induction
verifier, and produces more than twice correct proofs and about 35\% more
correct alarms than when analysing a large set of public available benchmarks.
\end{abstract}

\section{Introduction}
\label{sec:introduction}

Software model checking has experienced significant
progress in the last two decades, however, one major
bottleneck for its practical applications is scalability.
In particular, Bounded Model Checking (BMC) is a promising
approach to software verification~\cite{Biere:1999:SMC:646483.691738}, but its
application to verify and refute properties in large code bases is limited
by the resource requirements of the technique~\cite{svcomp2017}.
This happens when BMC techniques unwind all loops and recursive functions up to
their given maximum bound or completeness threshold~\cite{Kroening2011}, which
is typically infeasible for checking realistic programs.

In contrast, a variant called \textit{k}-induction
applied to unbounded programs uses BMC as a ``component'' to prove partial correctness~\cite{DBLP:books/daglib/0019162};
it has been successfully combined with continuously-refined invariants~\cite{Beyer15} and to prove that ANSI-C
programs do not contain data races~\cite{Kinductor,Donaldson10} or that
design-time constraints are respected~\cite{EenS03}. Additionally, \textit{k}-induction is a well-established
technique in hardware verification, where it is applied due to the transition relation
present in hardware designs~\cite{EenS03,GrosseLD09,Sheera00}. Although we can
prove partial correctness via induction without fully unwinding a program, state-of-the-art \textit{k}-induction
procedures still waste time and resources to falsify properties in programs since they unwind loops and
recursion up to the depth that exposes a bug.

Here we describe and evaluate a bidirectional
\textit{k}-induction (\textit{bkind}) algorithm, which is an extension of the original \textit{k}-induction~\cite{Sheera00}
 that improves its bug-finding capabilities, reducing the number of
iterations to find a property violation in half. In practice, the \textit{bkind}
algorithm performs a bidirectional search for bugs in the program state space
to quickly refute properties. Given the current knowledge in software model checking,
our extension has not previously been described or evaluated in the literature,
but we have already provided preliminary results of this approach on a limited
number of small benchmarks~\cite{GadelhaCN17}.
Similar techniques do exist, however, in other domains: Bischoff et
al.~\cite{BISCHOFF200533} describe a technique called ``target enlargement''
which combines binary decisions diagrams (BDDs) and Boolean Satisfiability (SAT)
solvers to reduce the time to find property
violations in hardware verification, and Bradley et al. introduced  ``property-directed reachability''
(or IC$3$) procedure for safety verification of systems~\cite{Bradley13} and have shown that
IC$3$ can scale on certain benchmarks, where \textit{k}-induction fails to succeed. Jovanovi\'{c} et
al.~\cite{Jovanovic:2016:PK:3077629.3077648} describe a technique called
``Property-Directed \textit{k}-induction'' to generate stronger invariants for
programs written in the SALLY input language.

In summary, this paper makes the following original contributions.
Firstly, we exploit the counterexamples generated by the $k$-induction proof
rule to derive new properties and feed them back to the BMC procedure.
Secondly, we combine an interval invariant generator and \textit{bkind} to
significantly improve the number of correct verification results. Lastly, our
experimental results show that \textit{bkind} can considerably reduce the
verification time compared to the na\"ive $k$-induction proof rule, since it
only requires half the number of steps to find a given safety property
violation in an unsafe program. Compared to other state-of-the-art
$k$-induction verifier (2LS), \textit{bkind} produces more than twice
correct proofs and about 35\% more correct alarms, when analysing a large set
of public available benchmarks.

\section{Na\"ive \textit{k}-Induction Proof Rule}
\label{kind:kind}

The first version of the \textit{k}-induction proof rule was proposed by
Sheeran et al.~\cite{Sheera00}. They used BMC algorithms to prove
correctness by induction. Consider a program $P$ with a loop and a safety
property $\phi(s)$. BMC algorithms can only show that no counterexample exists
for a $k$ loop unwindings but not that longer counterexamples
do not exist\footnote{Note that Craig interpolants can be used to exploit the
SAT/SMT solvers' ability to produce refutations, i.e., proofs that there is no
counterexample of depth less than or equal to
$k$~\cite{DBLP:reference/mc/McMillan18}.}. The
\textit{k}-induction proof rule tries to prove by induction that if $\phi$
holds for any given iteration through the loop then $\phi$ holds for the next
iteration. In particular, the base case tries to find a counterexample where
$\phi$ does not hold and the inductive step tries to prove that there exists no
counterexamples. {\it k}-induction extends induction by assuming the safety
property $k-1$ times before checking its satisfiability~\cite{Beyer15}, as
described in Eq.~\ref{eq:kind-nat}.
\begin{equation}\label{eq:kind-nat}
\footnotesize
 \left(\bigwedge_{i=0}^{k-1} \phi(i) \wedge \forall n :
\left(\left(\bigwedge_{i=0}^{k-1} \phi(n + i)\right) \Rightarrow \phi(n +
k)\right)\right) \Rightarrow \forall n: \phi(n).
\end{equation}

Since {\it k}-induction assumes the safety property $\phi$ more than once, a
less general case is checked after each iteration, thus it is more likely to
succeed~\cite{kind-principle}. In a previous work~\cite{MorseCNF13,Gadelha2015}, we
extended the \textit{k}-induction proof rule to check
program completeness in a separate step and defined it as
an iterative deepening algorithm~\cite{Russell:2003:AIM:773294}, consisting of
three independent checks: base case, forward condition and inductive step.

We describe the algorithms in this paper by assuming that a given
program $P$ under verification is a state transition system $M$. In $M$, a
state $s \in S$ is a tuple $(m,c)$, where $m$ is the state variable data, $c$
is the state constraint data. A predicate $init_P(s)$ denotes that $s$ is an
initial state, $tr_p(s_i, s_j) \in T$ is a transition relation from $s_i$ to
$s_{j}$, $\phi(s)$ is the formula encoding for states satisfying a safety
property, and $\psi(s)$ is the formula encoding for states satisfying a
completeness threshold~\cite{Kroening2011}, which is equal to the maximum number
of loop iterations occurring in $P$. For convenience, we define an error state
$\epsilon$, reachable if there is a property violation in the program
$P$. A counterexample $\pi^k$ is a sequence of states of length $k$ from the
initial state $s_1$ to $\epsilon$.

\

\begin{algorithm}[H]
\scriptsize
  \SetKwFunction{TheFn}{base\_case}
  \SetKwProg{Fn}{Function}{:}{}
  \Fn{\TheFn{$P$}}{
  \eIf{$init_P(s_1) \wedge \bigwedge^{j-1}_{i=1} tr_P(s_i, s_{i+1}) \wedge
\bigvee^{j}_{i=1} \neg \phi(s_i)$}{ \label{bc:cond}
    $\text{Let } \epsilon = s_i \text{ such that } \neg \phi(s_i)$\;
\label{bc:let}
    \KwRet$[s_1, \ldots, \epsilon]$\; \label{bc:cex}
  }{
    \KwRet$\emptyset$\;}
  }
  \caption{The base case.}
  \label{kind:fn-base-case}
\end{algorithm}

\begin{lemma}[Base case]\label{lemma-bc}
If the function \texttt{\upshape base\_case(P)} returns a sequence of states,
then the program is unsafe and the sequence of states is a counterexample.
\end{lemma}

\textit{Proof.} The execution path returned in line~\ref{bc:cex} of the
Algorithm~\ref{kind:fn-base-case} is a counterexample since it is an
execution path (ensured by the transition relation $tr_P$ in line~\ref{bc:cond})
that starts with the first state of the program (ensured by $init_P(s1)$ in
line~\ref{bc:cond}) and ends with an error state (in line~\ref{bc:let}); this
follows the definition of a counterexample. This is also a non-spurious counterexample
because the base case is a precise check: it encodes all reachable states up to
$k$ and checks for satisfiability. If the base case returns
a real counterexample then the program is unsafe.

\

\begin{algorithm}[H]
 \scriptsize
  \SetKwFunction{TheFn}{forward\_condition}
  \SetKwProg{Fn}{Function}{:}{}
  \Fn{\TheFn{$P$}}{
  \eIf{$init_P(s_1) \wedge \bigwedge^{j-1}_{i=1} tr_P(s_i, s_{i+1}) \wedge \neg
\psi(s_j)$}{ \label{fc:cond}
    \KwRet$[s_1, \ldots, s_j]$\;
  }{
    \KwRet$\emptyset$\;}
  }
  \caption{The forward condition.}
  \label{kind:fn-forward-cond}
\end{algorithm}

\begin{lemma}[Forward condition]\label{lemma-fc}
If the function \texttt{\upshape forward\_condition(P)} returns an empty
sequence of states then the program is safe.
\end{lemma}

\textit{Proof.} The forward condition checks if the completeness threshold was
reached in current unwound program $P$, i.e., all loops were completely
unwound. This is encoded as the completeness threshold property check in
line~\ref{fc:cond} of Algorithm~\ref{kind:fn-forward-cond}. In practice, these
checks are encoded as unwinding assertions and they check if the termination
condition of all loops are satisfiable for the current number of unwindings.
This step can prove partial correctness if the base case did not find any bug for
the current unwinding since no safety property is checked. We
guarantee this precedence in \textit{k}-induction by checking the
base case before checking the forward condition. We conclude that if no bug was
found by the base case and the completeness threshold holds for the current
number of unwindings, all states were explored and the program is safe.

\

\begin{algorithm}[H]
 \scriptsize
  \SetKwFunction{TheFn}{inductive\_step}
  \SetKwProg{Fn}{Function}{:}{}
  \Fn{\TheFn{$P$}}{
  \eIf{$\exists n \in \mathbb{N}^+. \bigwedge^{n + j - 1}_{i=n} (\phi(s_i)
\wedge tr_P(s_i, s_{i+1})) \wedge \neg \phi(s_{n+j})$}{  \label{is:cond}
    $\text{Let } \epsilon = s_{n+j} \text{ such that } \neg \phi(s_{n+j})$\;
\label{is:let}
    \KwRet$[s_n, \ldots, \epsilon]$\; \label{is:cex}
  }{
    \KwRet$\emptyset$\;}
  }
  \caption{The inductive step.}
  \label{kind:fn-inductive-step}
\end{algorithm}

\begin{lemma}[Inductive step]\label{lemma-is}
If the function \texttt{\upshape inductive\_step(P)} returns an empty
sequence of states, then the program is safe.
\end{lemma}

\textit{Proof.} Similarly to the forward condition check, the
program is safe up to $k$ loop unwindings because the base case did not find any
reachable error state. This is guaranteed in the \textit{k}-induction
proof rule by running the base case before the inductive step. The inductive
step then tries to find any counterexample of length $k$ in the state space by
first assuming that there was no property violation in $k-1$ iterations. The
inductive step over-approximates the state space so if no counterexample is
found then this is sufficient to prove that there is no reachable bug in the
program.

\

\begin{algorithm}[H]
 \scriptsize
  \SetKwFunction{TheFn}{kind}
  \SetKwProg{Fn}{Function}{:}{}
  \Fn{\TheFn{$P$, $k_{max}$, $k$}}{
  \lIf{$k > k_{max}$}{\KwRet{unknown}}\label{alg:kind-cond}
  \
  $P_k$ := unwinding$(P, k)$\;
  \
  $\pi$ := base\_case($P_k$)\; \label{alg:kind-bc1}
  \lIf{$\pi \neq \emptyset$}{\KwRet{$\pi$}} \label{alg:kind-bc2}
  \
  $\pi$ := forward\_condition($P_k$)\; \label{alg:kind-fc1}
  \lIf{$\pi = \emptyset $}{\KwRet{$\emptyset$}} \label{alg:kind-fc2}
  \
  $\pi$ := inductive\_step($P_k$)\; \label{alg:kind-is1}
  \lIf{$\pi = \emptyset $}{\KwRet{$\emptyset$}} \label{alg:kind-is2}
  \
  \KwRet{\texttt{\upshape kind}$(P, k_{max}, k + 1)$}\; \label{alg:kind-incr}
}
\caption{Na\"ive \textit{k}-induction.}
\label{alg:kind}
\end{algorithm}

\

The na\"ive version of the \textit{k}-induction proof rule (shown in
Algorithm~\ref{alg:kind}) tries to find a property violation or
to prove partial correctness for an increasing number of $k$ loop unwindings.
The pre-condition of the algorithm is $k=1$. The \texttt{unwinding} function
unwinds the program $P$, $k$ times; the function preserves the program
behaviour up to $k$ loop unwindings; if a bug in $P$ is reachable  in $k$
unwindings, it will be reachable in $P_k$. If it reaches a maximum number of
iterations $k_{max}$, the algorithm terminates with an \textit{unknown} answer.

\begin{theorem}[Soundness of the \textit{k}-induction proof rule]
\label{kind:sound}
If the k-induction proof rule returns: {\it (i)} a sequence of states $[s_i, \ldots, s_j]$: the program is unsafe and
the sequence is a non-spurious counterexample; {\it (ii)} $\emptyset$: the program is safe; {\it (iii)} \textit{unknown}: the program is safe up to $k_{max}$ iterations.
\end{theorem}

\textit{Proof.} The first item is ensured by Lemma~\ref{lemma-bc}, if there is
a property violation reachable after $k$ unwindings, the program is unsafe and
the algorithm terminates returning the counterexample (line~\ref{alg:kind-bc2}). The second item is ensured by Lemmas~\ref{lemma-fc} and~\ref{lemma-is}, if no
counterexample is found then the program is safe and an empty execution path is
returned in lines~\ref{alg:kind-fc2} and~\ref{alg:kind-is2}. Finally, the third item is
ensured in line~\ref{alg:kind-cond}, if the algorithm
reached the maximum number of defined iterations without terminating, an unknown
answer is given. We can then conclude that the \textit{k}-induction proof rule always terminates
either with a counterexample (if the program is unsafe), an empty execution
path (if the program is safe), or with an unknown answer otherwise.

\begin{theorem}[Partial completeness of the \textit{k}-induction
algorithm]\label{kind:completeness}
If $\exists~k: 1 \leq k \leq k_{max}$ such that the shortest counterexample is
$\pi^{k}_{min} = [s_1, \ldots, s_k]$ and $s_k = \epsilon$  then the
\textit{k}-induction proof rule will find the counterexample in at least $k$
iterations.
\end{theorem}

\textit{Proof.} This is ensured by always starting the \textit{k}-induction
algorithm as defined in Algorithm~\ref{alg:kind}  with one loop unwinding;
it always increments the number of loop unwindings by one
(line~\ref{alg:kind-incr}). Furthermore, if the program is unsafe, neither the
forward condition nor the inductive step will terminate the verification before
the counterexample is found. Also note that, if a property violation requires
zero loop unwindings (e.g., a property violation before a loop), the
\textit{k}-induction proof rule will still unwind the program once but
the base case will find the property violation since it checks
all states reachable with one loop unwinding (line~\ref{bc:cond}).

\subsection{Why is the \textit{k}-induction proof rule na\"ive?}

The inductive step assumption of all possible sequences of $k$
iterations is what makes the \textit{k}-induction proof rule na\"ive; these
sequences often include large unreachable regions of the state space. Safety
properties might not hold in these regions of the state space but they are
irrelevant for the safety of the program. For example, when verifying the safe
program in Fig.~\ref{fig:kind-example-safe} using \textit{k}-induction, the
inductive step will try to prove that the program is safe for all possible
values that both \texttt{input} and \texttt{s} variables can assume; this will
result in a series of spurious counterexamples since these variables
only assume a small range of values. In fact, the version of the
\textit{k}-induction proof rule as presented in Algorithm~\ref{alg:kind}
assumes that all counterexamples produced by the inductive step are
spurious, even if they are not.
\begin{figure}[!h]
\centering
  \begin{subfigure}{.46\textwidth}
\begin{lstlisting}[escapechar=^]
unsigned int s = 1;^\label{kind:example-ass0}^
int main() {
  while (1) {
    unsigned int input = __VERIFIER_nondet_int();
    if (input > 5) {^\label{kind:example-guard0}^
      return 0;
    } else if (input == 1 && s == 1) {^\label{kind:example-guard1}^
      s = 2;^\label{kind:example-ass1}^
    } else if (input == 2 && s == 2) {^\label{kind:example-guard2}^
      s = 3;^\label{kind:example-ass2}^
    } else if (input == 3 && s == 3) {^\label{kind:example-guard3}^
      s = 4;^\label{kind:example-ass3}^
    } else if (input == 4 && s == 4) {^\label{kind:example-guard4}^
      s = 5;^\label{kind:example-ass4}^
    } else if (input == 5 && s > 5) { ^\label{kind:example-guard5}^
      // unsatisfiable
      // property violation
      __VERIFIER_error(); ^\label{kind:example-assert}^
    }
  }
}\end{lstlisting}
\caption{Simplified safe program.}
\label{fig:kind-example-safe}
\end{subfigure}
\hfill
  \begin{subfigure}{.48\textwidth}
\begin{lstlisting}[escapechar=^]
unsigned int s = 1;^\label{kind:example-unsafe-ass0}^
int main() {
  while (1) {
    unsigned int input = __VERIFIER_nondet_int();
    if (input > 5) {^\label{kind:example-unsafe-guard0}^
      return 0;
    } else if (input == 1 && s == 1) {^\label{kind:example-unsafe-guard1}^
      s = 2;^\label{kind:example-unsafe-ass1}^
    } else if (input == 2 && s == 2) {^\label{kind:example-unsafe-guard2}^
      s = 3;^\label{kind:example-unsafe-ass2}^
    } else if (input == 3 && s == 3) {^\label{kind:example-unsafe-guard3}^
      s = 4;^\label{kind:example-unsafe-ass3}^
    } else if (input == 4 && s == 4) {^\label{kind:example-unsafe-guard4}^
      s = 5;^\label{kind:example-unsafe-ass4}^
    } else if (input == 5 && s >= 5) { ^\label{kind:example-unsafe-guard5}^
      // satisfiable
      // property violation
      __VERIFIER_error(); ^\label{kind:example-unsafe-assert}^
    }
  }
}\end{lstlisting}
\caption{Simplified unsafe program.}
\label{fig:kind-example-unsafe}
\end{subfigure}
\caption{Simplified illustrative examples extracted from SV-COMP'18 encoding an
event-condition-action (ECA) system~\cite{10.1007/978-3-662-43376-8_3}. The
program in Fig.~\ref{fig:kind-example-safe} is safe since the property
violation is unreachable while the program in Fig.~\ref{fig:kind-example-unsafe}
is unsafe since the property violation is reachable after at least $5$
iterations.}
\label{figure:kind-examples}
\end{figure}

Let us use an illustrative example to show the verification process using
\textit{k}-induction. First, consider the safe program in
Fig.~\ref{fig:kind-example-safe}. The property violation is reachable if the
transition condition $[input = 5 \wedge s > 5]$ holds and, since the
state space is over-approximated, there are several states that will satisfy
this condition. In this case, the \textit{k}-induction as defined in
Algorithm~\ref{alg:kind} will eventually reach the maximum number of loop
unwindings and terminate with an \textit{unknown} answer because the base case
will not find a property violation (the program is safe), the completeness
threshold will never be reached since the program contains an infinite loop and
the inductive step will keep finding spurious counterexamples.
Now, let us consider the unsafe program in Fig.~\ref{fig:kind-example-unsafe}.
The \textit{k}-induction as defined in Algorithm~\ref{alg:kind} will need at
least five iterations until a counterexample is found by the base case. During
these iterations, both the forward condition and the inductive step are
executed and any reasoning performed in these steps are discarded \textbf{but
what if the inductive step finds an actual partial counterexample?} This useful
information (a \textit{partial counterexample}) is ignored as all
counterexamples found by the inductive step are assumed to be spurious.

When using the na\"ive \textit{k}-induction to verify the programs in
Fig.~\ref{figure:kind-examples}, it will either produce an unknown result
or will discard useful information.

\section{Learning from Counterexamples to Bidirectionally Explore the State Space}
\label{kind:extensions}

The \textit{k}-induction proof rule can be applied to solve various verification problems~\cite{Kinductor,Donaldson10,EenS03}, but it can
be further improved by taking advantage of two important observations:
(1) \textit{partial counterexamples are ignored}: useful counterexamples may be
generated by the inductive step and they are ignored by the algorithm;
(2) \textit{unconstrained state space}: the inductive step may find spurious
counterexamples if the over-approximation is unconstrained. Several authors
address the later by generating program invariants to rule out unreachable
regions of the state space, either as a pre-processing step where invariants
are introduced in the program before~\cite{DBLP:conf/sbesc/RochaICB15,DBLP:conf/tacas/RochaRIC017} or
during the verification~\cite{Beyer15,10.1007/978-3-319-89963-3_24,Brain2015}.
Our algorithm is the first to address the former in the context of software
verification.

\subsection{Bidirectional bug-finding using \textit{k}-induction}
\label{kind:extension-bkind}

The \textit{bkind} algorithm extends the bug-finding capabilities of the
\textit{k}-induction proof rule by performing two alternating bug searches, one
forward (i.e., from the initial state $s_1$) and one backward (i.e., from any error
state $\epsilon$) and stopping if the forward search finds a state in a
counterexample produced by the backward search. Our proposed algorithm is
similar to the bidirectional search algorithm from the graph theory
field~\cite{DBLP:conf/aaai/SturtevantF18}. This new algorithm relies on two
checks from the \textit{k}-induction
proof rule to implement the searches. The base case is the forward search, since
it tries to find a counterexample $\pi^k = [s_1, \ldots, \epsilon]$, while
the inductive step is the backward search and tries to find any partial
counterexample $\pi^k = [s_i, \ldots, \epsilon]$. We shall refer to the base
case and inductive step as forward and backward searches, respectively.

To perform the forward search we
need to extend the base case as shown in Algorithm~\ref{kind:fn-new-base-case}.
First we define a new function \texttt{starts\_counterexample} that given a
state $s$ and a counterexample $\pi$, returns $true$ if $s \in \pi$
otherwise returns $false$; this function will be used to perform the
bidirectional search. The first condition in the new algorithm
(line~\ref{bkind:bc-old-cond}) is the same condition in the base case from the
original \textit{k}-induction proof rule and returns a counterexample if a bug
was found in $k$ iterations. The second (and new) condition
(line~\ref{bkind:bc-new-cond}) uses the function
\texttt{starts\_counterexample} to check if any of the states reachable by the
base case start the counterexample $\pi_{back}$ found by the backward search.
If this holds, the new base case function returns the execution path found by
the forward search concatenated with the counterexample found by the backward
search (line~\ref{nbc:concat}). The ``$\cdot$'' operator concatenates two
sequences. If no bug is found, the algorithm returns
an empty sequence.

\

\begin{algorithm}[H]
 \scriptsize
  \SetKwFunction{TheFn}{starts\_counterexample}
  \SetKwProg{Fn}{Function}{:}{}
  \Fn{\TheFn{$s$, $\pi$}}{
  \uIf{$\pi \neq \emptyset \wedge s \in \pi$}{
    \KwRet{true}\;
  }
  \Else{
    \KwRet{false}\;
  }
  }

  \

  \SetKwFunction{TheFn}{bkind\_base\_case}
  \SetKwProg{Fn}{Function}{:}{}
  \Fn{\TheFn{$P$, $\pi_{back}$}}{
  BC := $init_P(s_1) \wedge \bigwedge^{j-1}_{i=1} tr_P(s_i,
s_{i+1})$\;\label{nbc:init}

  \uIf{$\texttt{\upshape BC} \wedge \bigvee^{j}_{i=1} \neg \phi(s_i))$}{
\label{bkind:bc-old-cond}
    $\text{Let } \epsilon = s_i \text{ such that } \neg \phi(s_i)$\;
    \KwRet$[s_1, \ldots, \epsilon]$\;
  }
  \uElseIf{$\texttt{\upshape BC} \wedge \exists~i: 1 \leq i \leq k \wedge
\texttt{\upshape starts\_counterexample}(s_i,
\pi_{back})$}{\label{bkind:bc-new-cond}
    \KwRet$[s_1, \ldots, s_i-1] \cdot \pi_{back}$\; \label{nbc:concat}
  }
  \Else{
    \KwRet$\emptyset$\;}
  }
  \caption{The base case used in the \textit{bkind} algorithm.}
  \label{kind:fn-new-base-case}
\end{algorithm}

\begin{lemma}[Base case]\label{lemma-nbc}
If Algorithm~\ref{kind:fn-new-base-case} returns a sequence of states
$[s_i,\ldots, s_j]$, this is a non-spurious counterexample.
\end{lemma}

\textit{Proof.} The first condition in the new base case
(line~\ref{bkind:bc-old-cond}) is identical to the condition in the original
base case and Lemma~\ref{lemma-bc} ensures that this is a real counterexample.
We only need to prove that the execution path returned in line~\ref{nbc:concat}
is a counterexample. The returned sequence is a counterexample since it is a
concatenation of an execution path starting from the initial state in the state
space with a counterexample. We know that the sequence of states in the
concatenation is an execution path because of the transition relation and
$init_P(s_1)$ ensures that it starts from the initial state in the state space
(both in line~\ref{nbc:init}). Lemma~\ref{partial-cex} guarantees that
$\pi_{back}$ is a partial counterexample. Finally, this is a non-spurious
counterexample because of the partial order property of the state space: this is
sufficient to allow the concatenation of the execution path and the
counterexample.

\

\begin{algorithm}[H]
\scriptsize
  \SetKwFunction{TheFn}{bkind}
  \SetKwProg{Fn}{Function}{:}{}
  \Fn{\TheFn{$P$, $k_{max}$, $k$, $\pi_{back}$}}{
  \lIf{$k > k_{max}$}{\KwRet{unknown}}\label{alg:bkind-cond}
  \
  $P_k$ := unwinding$(P, k)$\;
  \
  $\pi$ := bkind\_base\_case($P_k, \pi_{back}$)\; \label{alg:bkind-bc1}
  \lIf{$\pi \neq \emptyset$}{\KwRet{$\pi$}} \label{alg:bkind-bc2}
  \
  $\pi$ := forward\_condition($P_k$)\; \label{alg:bkind-fc1}
  \lIf{$\pi = \emptyset $}{\KwRet{$\emptyset$}} \label{alg:bkind-fc2}
  \
  $\pi$ := inductive\_step($P_k$)\; \label{alg:bkind-is1}
  \lIf{$\pi = \emptyset $}{\KwRet{$\emptyset$}} \label{alg:bkind-is2}
  \
  \KwRet{\texttt{\upshape bkind}$(P, k_{max}, k + 1, \pi)$}\;
\label{alg:bkind-incr}
}
\caption{The \textit{bkind} algorithm.}
\label{alg:bkind}
\end{algorithm}

\

The \textit{bkind} algorithm is shown in Algorithm~\ref{alg:bkind}. Similarly to
the na\"ive \textit{k}-induction algorithm, the \textit{bkind} algorithm tries to either
find a property violation or to prove partial correctness for an increasing number of
$k$ unwindings. The pre-conditions of the algorithm are $\pi=\emptyset$ and
$k=1$. If it reaches a maximum number of iterations $k_{max}$, the algorithm
terminates with an \textit{unknown} answer. The novel contribution in the new
\textit{bkind} algorithm is the bidirectional bug-finding technique. We use
the counterexample produced by the backward search in the previous iteration,
and check if it is reachable by the forward search in the next iteration
(line~\ref{alg:bkind-bc1}).

\begin{lemma}[Partial counterexample from the inductive step]\label{partial-cex}
If Algorithm~\ref{kind:fn-inductive-step} returns a sequence of states
$[s_i,\ldots, s_j]$, this is a partial counterexample of length $k$.
\end{lemma}
\textit{Proof.} This is a partial counterexample because it is an
execution path (the transition relation in line~\ref{is:cond} ensures that) and
the last state in the path is an error state (ensured in line~\ref{is:let}).
Finally, the counterexample has length $k$
because the inductive step always tries to find a counterexample of length $k$;
this is performed by checking if the property violation is reachable in $k$
iterations, assuming that it holds for $k-1$ iterations.

\begin{theorem}[Partial completeness of the \textit{bkind}
algorithm]\label{kind:halving}
If $\exists~k: 1 \leq k \leq k_{max}$ such that the shortest counterexample is
$\pi^{k}_{min} = [s_1, \ldots, s_k]$ and $s_k = \epsilon$  then the
\textit{k}-induction proof rule will find the counterexample in at least
$\lfloor\frac{k}{2}\rfloor+1$ iterations.
\end{theorem}

\textit{Proof.} In order to prove this theorem, we assume that the  inductive
step always returns the same non-spurious partial counterexample for every $k$;
we will show how to give partial guarantees to this assumption in
Sec.~\ref{kind:extension-interval}.

First, we show that no more than $\lfloor\frac{k}{2}\rfloor+1$ iterations are
required to find a property violation. By contradiction, assume that the number
of iterations required to find the property violation is greater than
$\lfloor\frac{k}{2}\rfloor+1$.

Let us assume a $\pi_{min}^k$ where $k$ is even. In the iteration $\frac{k}{2}$,
the new base case will have explored all states up to $\frac{k}{2}$ and the
inductive step will have provided a partial counterexample
$\pi_{back}^{\frac{k}{2}}$. In the iteration $\frac{k}{2}+1$, the new base case
will not reach any state in $\pi_{back}^{\frac{k}{2}}$ if either the
counterexample $\pi_{back}$ is spurious (i.e., contradicting our initial assumption)
or the counterexample $\pi_{back}$ is not spurious and there is at least
one state $s_u \in \pi^k_{min}$ such that $\pi_{min}^k = [s_1,
\ldots, s_{\frac{k}{2} + 1}] \cdot [s_u, \ldots ] \cdot \pi^{\frac{k}{2} +
1}_{back}$, which has a length greater than $k$ contradicting our assumption
about the length of $\pi^k_{min}$.

Now, assume a $\pi_{min}^k$ where $k$ is odd. The proof is similar to the one
where $k$ is even, except that we consider the sequences at iteration
$\frac{k-1}{2}$. If the counterexample is not found in the iteration
$\frac{k-1}{2}+1$, then either the counterexample is spurious or a
greater number of iterations is required to find the property violation,
contradicting our initial assumptions.
We then generalize and conclude that no more than $\lfloor\frac{k}{2}\rfloor+1$
iterations are required to find a property violation.

Now we show that at least $\lfloor\frac{k}{2}\rfloor+1$ iterations are
required to find a property violation. By contradiction, assume
that the number of iterations required to find the property violation is
less than $\lfloor\frac{k}{2}\rfloor+1$. This means that either there
exists a smaller counterexample that was not found the base case, which violates
Lemma~\ref{lemma-nbc}, or there is a state $s_v \in \pi^k_{min}$
such that $\pi^{k}_{min} = [s_1, \ldots, s_v] \cdot [s_{v+1}, \ldots, s_k]$
which has a length smaller than $k$ contradicting our assumption about the
length of $\pi^k_{min}$. Given that at least $\lfloor\frac{k}{2}\rfloor+1$ iterations are needed to find
the property violation and no more than $\lfloor\frac{k}{2}\rfloor+1$
iterations are needed to find the property violation, we can conclude that the
\textit{bkind} algorithm will find a counterexample $\pi_{min}^k$ in
exactly $\lfloor\frac{k}{2}\rfloor+1$ iterations.

\begin{theorem}[Soundness of the \textit{bkind} algorithm]
\label{bkind:sound}
If the bkind algorithm returns: {\it (i)} a sequence of states $[s_i, \ldots, s_j]$: the program is unsafe and
the sequence is a non-spurious counterexample; {\it (ii)} $\emptyset$: the program is safe; {\it (iii)} \textit{unknown}: the program is safe up to $k_{max}$ iterations.
\end{theorem}

\textit{Proof.} The proof follows the same proof as the
Theorem~\ref{kind:completeness}, except that the first item is ensured by
Lemma~\ref{lemma-nbc} instead of Lemma~\ref{lemma-bc}.

\subsection{Constraint Generation Using Interval Analysis}
\label{kind:extension-interval}

\begin{figure}[!h]
  \centering
  \includegraphics[width=0.85\textwidth]{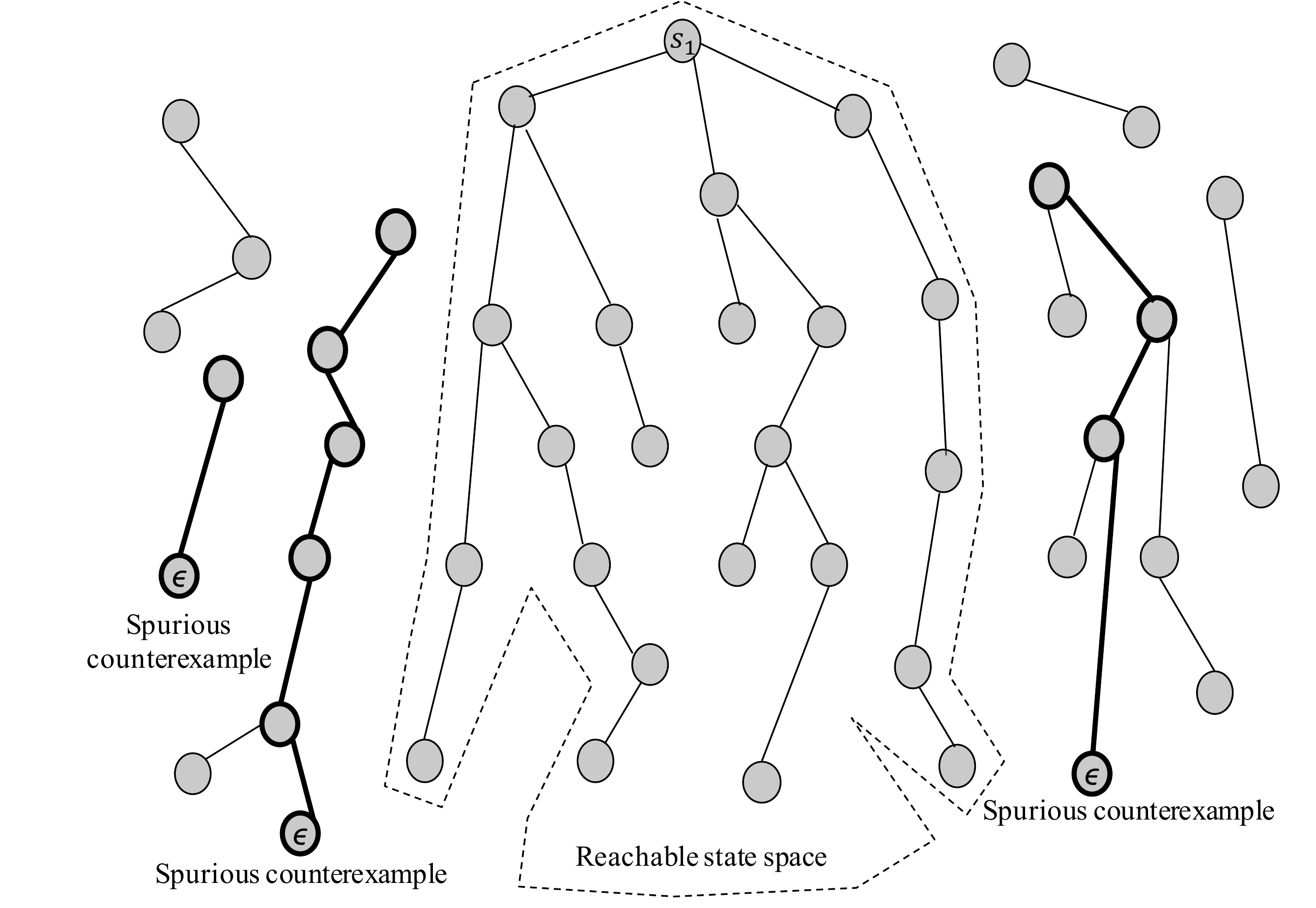}
  \caption{Visual representation of the state space of a program, where the
states
inside the dashed line are the reachable states and the ones outside are
unreachable states. An unconstrained over-approximation of the program assumes
that all the states are reachable which might contain spurious counterexamples
(counterexamples that lead to unreachable error states). An invariant is a
filter of states: a strong enough invariant will remove unreachable error
states from reasoning, allowing the inductive step to prove the program
correctness or to find non-spurious partial counterexamples.}
  \label{figure:kind-inv}
\end{figure}

Here we 
use invariants to
constraint the state and rule out unreachable states evaluated by the inductive
step~\cite{Beyer15,DBLP:conf/tacas/RochaRIC017,10.1007/978-3-319-89963-3_24,Brain2015}.
Fig.~\ref{figure:kind-inv} shows an
example of the usage of invariants (dashed line) to constraint the state space.
The invariants reduce the number of states explored by the backward search by
constraining the over-approximation.
Similarly to Rocha et al.~\cite{DBLP:conf/tacas/RochaRIC017}, we perform a static program analysis
prior to loop unwinding and estimate the intervals that a variable can
assume. In contrast to Rocha et al., we do not rely on external tools
and implement the invariant generation as a pre-processing step of the
verification. In particular, we use the abstract interpretation component from
CProver~\cite{cprover-manual}. The invariant generation algorithm
uses an abstract domain based on expressions over intervals, such that every
constraint $c$ in a state $s$ is a map $var \rightarrow 2^{128} \times 2^{128}$
that maps interval constraints to every variable $var$.

In order to use the invariants, we need
to extend the inductive step as shown in
Algorithm~\ref{kind:fn-new-inductive-step}. The algorithm is similar to the
inductive step as defined in Algorithm~\ref{kind:fn-inductive-step}, but it now
takes an extra argument: a set of invariants $\varphi$. These invariants will
constraint the state space and filter unreachable states from the inductive
step check, thus reducing the number of spurious path that might be explored.

\

\begin{algorithm}[H]
 \scriptsize
  \SetKwFunction{TheFn}{inductive\_step\_invariants}
  \SetKwProg{Fn}{Function}{:}{}
  \Fn{\TheFn{$P$, $\varphi$}}{
  \eIf{$\exists n \in \mathbb{N}^+.~\varphi(s_n) \bigwedge^{n + j - 1}_{i=n}
(\phi(s_i) \wedge tr_P(s_i, s_{i+1})) \wedge \neg \phi(s_{n+j})$}{
\label{iis:cond} $\text{Let } \epsilon = s_{n+j} \text{ such that } \neg
\phi(s_{n+j})$\;
\label{iis:let}
    \KwRet$[s_n, \ldots, \epsilon]$\; \label{iis:cex}
  }{
    \KwRet$\emptyset$\;}
  }
  \caption{The new inductive step with invariants.}
  \label{kind:fn-new-inductive-step}
\end{algorithm}

\begin{lemma}[Inductive step with invariants]\label{lemma-iis}
If \texttt{\upshape inductive\_step\_invariants(P)} returns an empty sequence
of states the program is safe.
\end{lemma}

\textit{Proof.} This follows the same reasoning of the Lemma~\ref{lemma-is}.
The program is safe up to $k$ iterations because the base case did
not find any property violation and that the inductive step over-approximates
the state space when it tries to find a property violation. The new inductive
step with invariants will constraint the over-approximation to be closer to the
reachable state space of the program.
Lemma~\ref{lemma-iis} ensures that the \textit{k}-induction proof rule
can use the new inductive step with invariants. Theorem~\ref{kind:sound}
must change if invariants are used to prove partial correctness;
Lemmas~\ref{lemma-fc} and~\ref{lemma-iis} ensure the theorem is sound.

\begin{lemma}[Partial counterexample from the inductive step with
invariants]\label{partial-cex-iis}
If Algorithm~\ref{kind:fn-new-inductive-step} returns a sequence of states
$[s_i,\ldots, s_j]$, this is a partial counterexample of length $k$.
\end{lemma}

\textit{Proof.} This follows the same reasoning of the Lemma~\ref{partial-cex}.
We know this is a counterexample because of the sequence of transitions defined
by $tr_P$ and that the last state is an error state. Again, the invariants
here will only constraint the state space so the over-approximation is closer to
the set of reachable states of the program.
The Lemma~\ref{partial-cex-iis} is defined so the new inductive step with
invariants can be used with the \textit{bkind} algorithm.
Theorem~\ref{bkind:sound} needs to be changed if invariants are used to prove
correctness to use Lemma~\ref{partial-cex-iis} instead of Lemma~\ref{lemma-is}.

\subsection{Why is the \textit{bkind} algorithm more efficient than
\textit{k}-induction?}
\label{kind:bkind-example}

First, consider that we wish to verify the safe program in
Fig.~\ref{fig:kind-example-safe} using the \textit{bkind} algorithm. The
state transition system is analyzed and the following intervals are estimated
based on the assignments: $\varphi = (input \geq 0, \: input \leq
\texttt{UINT\_MAX}, \: s \geq 1, \: s
\leq 5)$. The invariants are introduced in the program and are sufficient to
prove that the program is safe safe with two loop unwindings: there will be no
counterexample of size two that leads to a property violation.
Now, consider that we wish to verify the unsafe program in
Fig.~\ref{fig:kind-example-unsafe}. Here, the same set of constraints are
introduced in the program but now the inductive step will find a counterexample
that satisfies $\footnotesize input \geq 0 \wedge input \leq \texttt{UINT\_MAX}
\wedge s \geq 1 \wedge s \leq 5 \wedge input = 5 \wedge s \geq 5$, which is $
input = 5 \wedge s = 5$.
%
%
This is the program state prior to the error state; the reachability of this
state is introduced in the program as a new property and checked in the base
case. This is then extended further back for every loop iteration, effectively
performing the backward search. In conclusion, \textit{bkind} can correctly
verify both programs in Fig.~\ref{figure:kind-examples}: the program in
Fig.~\ref{fig:kind-example-safe} can be proven to be safe and the program in
Fig.~\ref{fig:kind-example-unsafe} requires fewer number of steps to find the
property violation.

\section{Experimental Evaluation}
\label{kind:results}

The experimental evaluation of the \textit{bkind} algorithm and the invariant
generation in our software model checker (ESBMC~\cite{CordeiroFM12,esbmc2018}) consists of three parts. In
Sec.~\ref{bkind-benchmarks-description}, we describe the experimental objectives and present the benchmarks used
to evaluate the \textit{bkind} algorithm. In Sec.~\ref{bkind-esbmc} we compare
our \textit{bkind}
algorithm and the invariant generation with the \textit{na\"ive}
\textit{k}-induction, while in Sec.~\ref{bkind-2ls} we compare the
\textit{bkind} algorithm with invariants against another state-of-the-art BMC
tool that uses \textit{k}-induction and invariant generation to verify ANSI-C
programs, 2LS~\cite{10.1007/978-3-319-89963-3_24}. The tools are compared in terms of number of refuted bugs and
verification time. We provide a virtual machine with all the binaries and scripts to
reproduce our results in \footnotesize{\url{www.esbmc.org}}.

\subsection{Experimental Objectives and Setup}
\label{bkind-benchmarks-description}

Our experimental evaluation aims to answer three research questions:
\begin{enumerate}
\item[RQ1] \textbf{(soundness)} Does our approach provide correct
results?
\item[RQ2] \textbf{(performance I)} Does our approach improve results compared
to the na\"ive \textit{k}-induction?
\item[RQ3] \textbf{(performance II)} How does our approach compare against
other \textit{k}-induction verifiers?
\end{enumerate}

We use $5591$ benchmarks from SV-COMP'18 to evaluate the algorithms described
in this paper. The benchmarks were extracted from the subcategories
\textit{Arrays}, \textit{BitVectors}, \textit{ControlFlow}, \textit{ECA},
\textit{Floats}, \textit{Heap}, \textit{Loops}, \textit{ProductLines},
\textit{Sequentialized} and \textit{Systems\_DeviceDriversLinux64}. The
remaining categories were excluded because they use features that our
\textit{k}-induction does not support (e.g., termination, recursion, and concurrency).
When verifying those programs, ESBMC disables the inductive step
and uses only the base case and the forward condition, thus they are not
included here. Out of the $5591$ benchmarks, $4134$ are safe while
$1457$ are unsafe programs.

All experiments were conducted on IRIDIS4, the supercomputer from the University
of Southampton~\cite{iridisSoton}. The computer nodes used are equipped with
Intel
Sandybridge processors running at $2.6$GHz and $24$GB of RAM. We used Boolector
as the SMT backend for all the verification tasks.
For each benchmark, we set time and memory limits of $900$ seconds and $15$GB,
respectively, as per the competition definitions. 
Finally, given the large amount of data involved in the experiments, we used
four
groups to present the results: \textit{Correct proofs} is the
number of correct positive results (i.e., the tool reports SAFE correctly),
\textit{Correct alarms} is the number of correct negative results (i.e., the
tool reports UNSAFE correctly), \textit{Incorrect proofs} is the number of
incorrect positive results (i.e., the tool reports SAFE incorrectly),
\textit{Incorrect alarms} is the number of incorrect false results (i.e.,
the tool reports UNSAFE incorrectly).

\subsection{Comparison of \textit{k}-induction-based approaches}
\label{bkind-esbmc}

Here, we evaluate five different \textit{k}-induction proof rules: 
``original na\"ive \textit{k}-induction'' (the first version implemented
in ESBMC~\cite{Gadelha2015}), ``na\"ive \textit{k}-induction'' (the
\textit{k}-induction proof rule described in Sec.~\ref{kind:kind}),
``na\"ive \textit{k}-induction + invariants'' (the \textit{k}-induction
algorithm described in Sec.~\ref{kind:kind} and the invariants described in
Sec.~\ref{kind:extension-interval}), ``\textit{bkind}'' (the \textit{bkind}
algorithm described in Sec.~\ref{kind:extension-bkind}) and ``\textit{bkind}
+ invariants'' (the \textit{bkind} algorithm described in
Sec.~\ref{kind:extension-bkind} and the invariants described in
Sec.~\ref{kind:extension-interval}).
\begin{figure}[!h]
\centering
\includegraphics[width=1\textwidth]{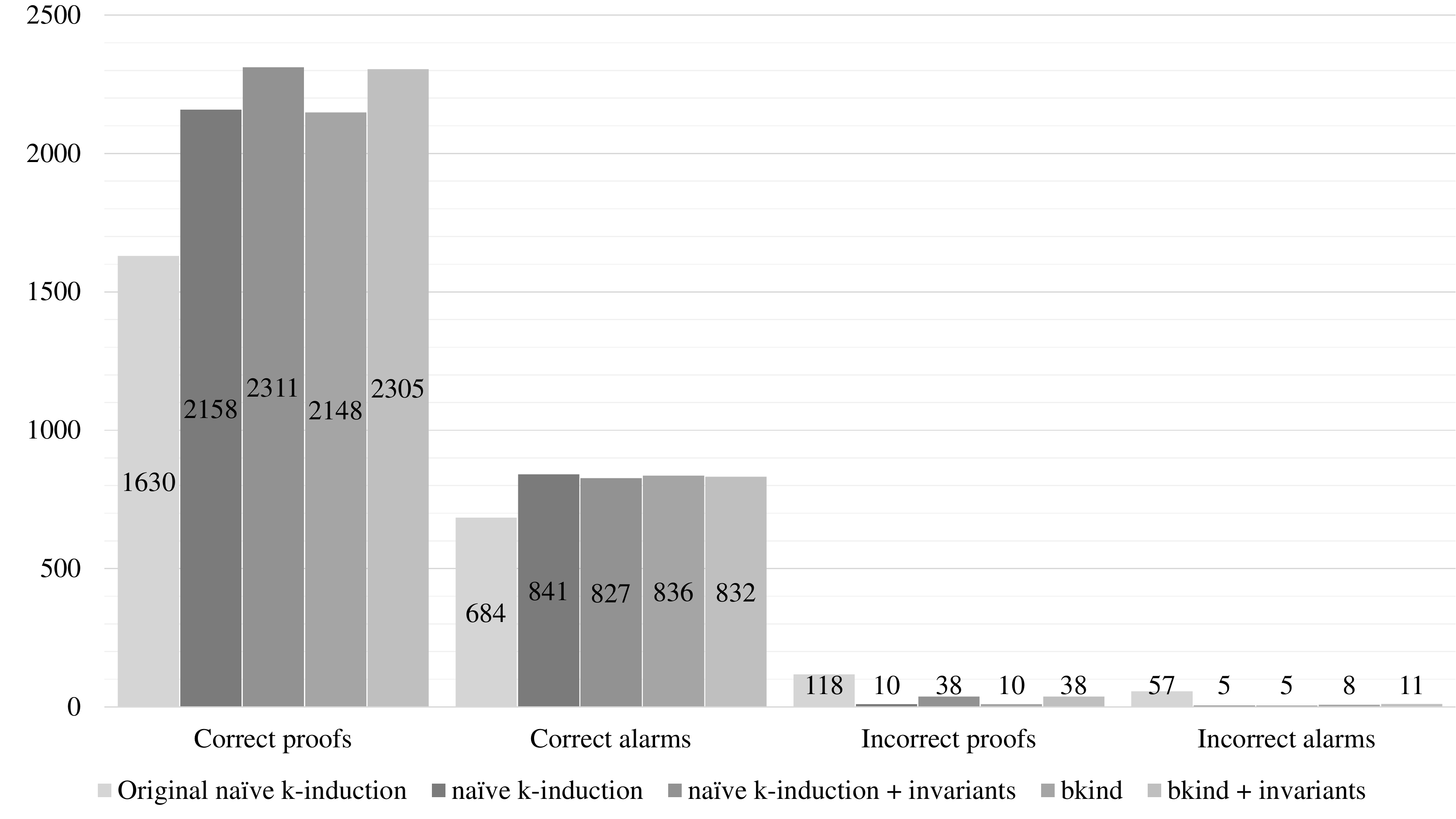}
\caption{Results of the \textit{k}-induction-based algorithms in ESBMC for all
SV-COMP'18 benchmarks with different configurations.}
\label{figure:kind}
\end{figure}

Fig.~\ref{figure:kind} shows the results of using the
\textit{k}-induction-based approaches to verify the
benchmarks from SV-COMP. First, let us compare the results of the original
na\"ive \textit{k}-induction~\cite{Gadelha2015} and the current na\"ive
\textit{k}-induction: the number of
correct proofs and correct alarms increased by 25\% and 20\%, respectively,
while the number of incorrect proofs and incorrect alarms decreased by 91\% and
92\%, respectively. Most of the wrong results in the original
\textit{k}-induction came from the fact the original algorithm (1) could
not reason about early loop exits (e.g., a \texttt{break} inside a loop) and (2)
would assume wrong safety conditions in the inductive step due to implementation bugs.
Furthermore, the original \textit{k}-induction did not support floating-point
encoding which resulted in $45$ incorrect alarms in the \textit{Floats}
category and did not have the clang frontend, thus it could not verify about 500
benchmarks due to parsing errors.


Now let us compare the current \textit{k}-induction proof rule against the new
\textit{bkind} algorithm (with and without invariants). First, we notice that
the invariants increase the number of correct proofs for both the
\textit{k}-induction and \textit{bkind} in about 7\%. This, however, comes at
a cost: due to bugs in our implementation, the number of incorrect proofs are
almost 4 times higher when invariants are used in combination with the
algorithms (from 10 to 38). In
particular, our algorithm does not track intervals of variables changed through
pointers and neither if the intervals are defined in terms of other variables.
The number of
incorrect results, however, is still low: we only report incorrect proofs in
about 2.5\% of the $1457$ incorrect benchmarks.

Note that all the approaches report similar correct alarms with or
without invariants; this is expected in the \textit{k}-induction proof rule
since the invariants are supposed to only improve the correctness proof. The
\textit{bkind} verification results could have found a larger number of bugs
than the \textit{k}-induction proof rule. Indeed, \textit{bkind}
finds bugs in benchmarks that could not be found by the \textit{k}-induction
algorithm but in the end it reported a slightly fewer number of correct alarms
and a larger number of incorrect alarms. An in-depth analysis of the wrong
results showed that (1) when the invariants are incorrect, the \textit{bkind}
algorithm ends up finding an incorrect counterexample and (2) when the program
contains arrays, the algorithm ends up generating incomplete partial
counterexamples, which also lead to incorrect alarms. Despite the number of
incorrect results, however, the wrong alarms only amount to 0.1\% of the $4134$
correct benchmarks analysed by the \textit{k}-induction approaches.
These numbers allow us to partially affirm our research question RQ1: the new
\textit{bkind} algorithm provides correct results for a large set of benchmarks.
There are some programs where \textit{bkind} will provide incorrect
results but it is due to bugs in our implementation.


The total verification time is $1,253,015$s for ``original na\"ive
\textit{k}-induction'', $1,654,147$s for  ``na\"ive
\textit{k}-induction'', $1,522,851$s for ``na\"ive \textit{k}-induction
+ invariants'', $1,664,357$s for ``\textit{bkind}'', and $1,504,208$s
for ``\textit{bkind} + invariants''. First, let us evaluate the original
and the current na\"ive \textit{k}-induction: the original one is 25\% faster
the current \textit{k}-induction; this can be easily explained due to the
limitations in the program: a number of benchmarks are not parsed by the tool
and the greater number of incorrect results allow the original
\textit{k}-induction to finish the analysis faster.
%
Regarding the \textit{k}-induction and \textit{bkind} without invariants, the
latter is slightly slower (0.6\%): this is expected since the inductive step is
most likely to find spurious counterexamples. The slowdown in
\textit{bkind} is the impact of introducing spurious verification conditions
and it is negligible.

The results when invariants are used, however, are much better. The
verification time decreases considerably in these benchmarks, making both
algorithms 10\% faster. The na\"ive \textit{k}-induction algorithm takes
$1,654,147$s to verify all the programs which is equivalent to $19.1$
days
of continuous processing, while the \textit{bkind} algorithm with invariants
takes $1,504,208$s or $17.4$ days. The \textit{bkind} algorithm with invariants
speeds up the verification by almost two full days in our experiments.
These results allow us to affirm our RQ2: the \textit{bkind}
algorithm with invariants improves the performance over the
na\"ive \textit{k}-induction by giving more correct results in less time.

\subsection{Comparison to a state-of-the-art \textit{k}-induction verifier 2LS}
\label{bkind-2ls}

\begin{figure}[!h]
\centering
\includegraphics[width=1\textwidth]{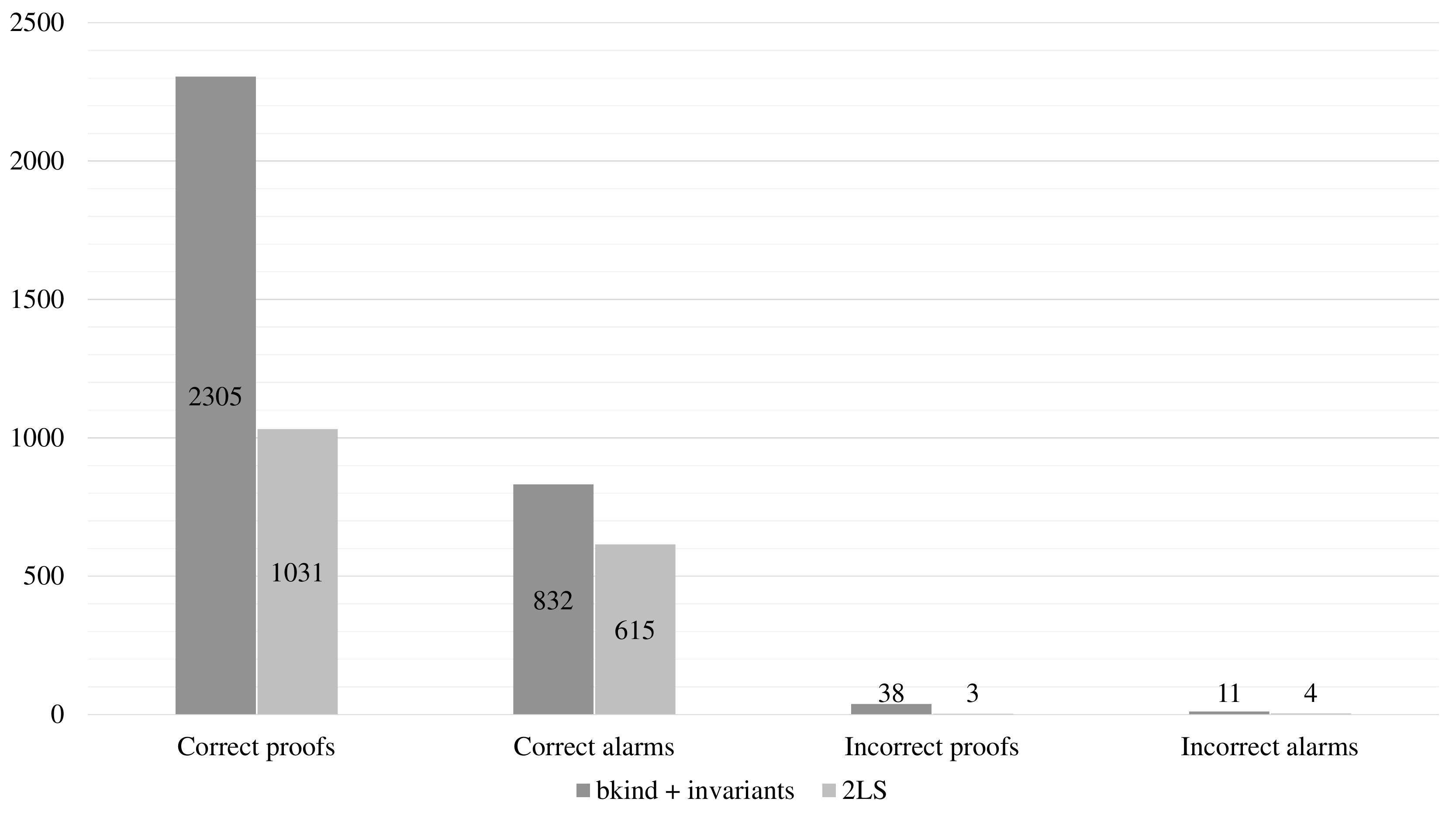}
\caption{Results of ESBMC with \textit{bkind} algorithm with invariants and 2LS.}
\label{figure:2ls}
\end{figure}

We now compare the \textit{bkind} algorithm with invariants against 2LS $v0.6.0$,
another state-of-the-art bounded model checker with support for
\textit{k}-induction.
In particular, 2LS uses the \textit{kIkI}
algorithm~\cite{10.1007/978-3-319-89963-3_24} and combines the
\textit{k}-induction proof rule with continuous invariant generation. We used
the same configuration from SV-COMP'18 in which 2LS is configured to generated
interval constraints similar to the ones generated by ESBMC.

As shown in Fig.~\ref{figure:2ls}, ESBMC with the \textit{bkind} algorithm and
invariants produces more than two times the number of correct proofs and about
35\% more correct alarms compared to 2LS when
analysing the same set of benchmarks. An in-depth analysis of the results show
that 2LS aborts the verification of various benchmarks in the
\textit{Systems\_DeviceDriversLinux64} category with the message
``Irreducible control flow not supported''. Alternatively, ESBMC is
able to prove the correctness of $1249$ benchmarks in this category, greatly
improving our results.
2LS provides much fewer incorrect results when compared to ESBMC in this set of
benchmarks. 2LS always had a strong focus in invariant generation since its
first version; their last version in SV-COMP18 extended it even further by
introducing invariant generation for termination proofs and pointer safety. The
invariant generation in ESBMC is still in its first version
and needs improvements.


The total verification time is $1,266,890$s for 2LS and $1,504,208$s for ESBMC
with ``\textit{bkind} + invariants''.
Here, 2LS is about 15\% faster than ESBMC even if we do not consider the
\textit{Systems\_DeviceDriversLinux64} category where 2LS would abort early in
the verification. In particular, 2LS was more than 10 times faster in two
categories, \textit{Arrays} and \textit{Heap}, most likely due to the stronger
invariants generated by the tool which allowed it to prove correctness faster;
ESBMC would simply run out of time in a large number of benchmarks in these
categories.
ESBMC with \textit{bkind} and invariants is not the fastest verification tool
using \textit{k}-induction but it is the one with the highest number of correct
results. These results allow us to answer the RQ3: \textit{our novel contribution is a
improvement over the state-of-the-art verification using
\textit{k}-induction, which can report more correct results for a large number of
different benchmarks than other existing approaches}.

\section{Related Work}
\label{kind:related}

The \textit{k}-induction method is gaining popularity
in the software verification community. Donaldson {\it et~al.}\ described a
verification tool called Scratch to detect data races during
Direct Memory Access (DMA) in the CELL BE processor from
IBM~\cite{Donaldson10}, using \textit{k}-induction. Properties are automatically
inserted in the program to model the behaviour of
the memory control-flow and the algorithm tries to find violation of those
properties or prove that they hold indefinitely. The method also requires the code to be manually
annotated with loop invariants, whereas our approach automatically generates
and adds them to the program. Finally, the tool is able to prove the absence of
data races, but it is restricted to verify that
specific class of problems for a particular type of hardware, while our
approach is evaluated over a more general class of programs.

Donaldson further described two tools for proving
correctness of programs: K-Boogie and K-Inductor~\cite{Kinductor}. The former
is an extension of the Boogie language, aimed to prove correctness (using
\textit{k}-induction) of programs written in a number of languages (e.g., Boogie and
Spec), while the latter is a BMC tool for C programs. Both K-Boogie and K-Inductor
use a \textit{k}-induction proof rule; the completeness threshold is
not separately checked and relies only on the inductive step to prove
correctness. Their \textit{k}-induction has a pre-processing step, but
while we introduce invariants during the pre-processing, their approach
removes all nested loops leaving only non-nested loops. They compare the
results of K-inductor with Scratch and show that the new approach maintains the
same number of correctly verified programs while being faster.
Similar to the prior work~\cite{Donaldson10}, the programs need to
be manually changed to add loop invariants while we do
it automatically.

Mal{\'i}k et al.~\cite{10.1007/978-3-319-89963-3_24} describe 2LS, a C/C++
SAT-based BMC. 2LS is a tool developed using the CProver
framework~\cite{Clarke04} and combines a \textit{k}-induction proof rule
with abstract interpretation (AI). As CBMC~\cite{Clarke04}, 2LS uses SAT solvers but instead of a
fixed unwind approach, 2LS uses an incremental BMC approach,
where it first checks for property violations for a given bound, then tries to
generate (and refine) invariants using AI and then builds
a proof using \textit{k}-induction. Their
\textit{k}-induction, called \textit{kIkI}~\cite{Brain2015}, is
similar to the one implemented in ESBMC, but adds an extra step to generate
and refine invariants. In contrast to our invariant generation that
only supports interval domains, 2LS supports several abstract domains for
numerical values and a shape domain for pointers. 2LS offers approaches to
prove non-termination, while ESBMC has no algorithm to prove non-termination and
can only prove termination by checking the unwinding assertions.

Bischoff {\it et al.}~\cite{BISCHOFF200533} propose a methodology to use BDDs and
SAT solvers for the verification of programs in a bidirectional form similar
to our \textit{bkind} algorithm. In their work, they refer to the technique as
\textit{target enlargement}: the property violation is ``enlarged'' by
checking if the states around the property violation are reachable. The BDDs
are responsible for the
target enlargement, collecting the under-approximate reachable state sets,
followed by the SAT-based verification with the newly computed sets. They
implemented the technique in the Intel BOolean VErifier and showed that
the verification time of a set of public benchmarks was up to five times
smaller. Compared to this work, we only use \textit{k}-induction and SMT
solvers; the inductive step in the \textit{k}-induction is
responsible for enlarging the target and the SMT solver checks for
satisfiability.

Jovanovi\'{c} {\it et al.}~\cite{Jovanovic:2016:PK:3077629.3077648} present a
reformulation of IC3, separating the reachability checking from the inductive
reasoning. They further replace the regular induction proof rule by the
\textit{k}-induction and show that it provides more concise
invariants. The authors implemented the algorithm in the SALLY model checker
using Yices2 to do the forward search and MathSAT5 to do the backward search.
They showed that the new algorithm can solve a number of real-world
benchmarks at least as fast as other approaches. Compared to this work, our
 \textit{bkind} uses consecutive BMC calls to find a
solution. We implement our approach independent of solvers and it can be
used with any SMT solver supported by ESBMC; both searches are
done with the same solver.

\section{Conclusions}
\label{kind:conclusions}

We have described the \textit{k}-induction proof rule and
a novel contribution that extended its bug-finding capabilities.
The new algorithm, called bidirectional {\it k}-induction or \textit{bkind}, was implemented in ESBMC
and evaluated in a large set of benchmarks.
\textit{k}-induction is a powerful verification technique implemented in
several different tools and was successfully used to verify a large number of
different programs and properties.
Here, we proposed and evaluated a novel way to exploit the
\textit{k}-induction proof rule, where useful information
can be extracted from the various checks in the algorithm and can be used to
improve the results of the algorithm.

In particular, the \textit{bkind} algorithm uses information extracted from the
inductive step to shorten the number of steps required to find a property
violation; with strong enough invariants the \textit{bkind} algorithm requires
roughly half of the number of loop unwindings a BMC algorithm requires to
find a property violation. We have implemented an interval invariant generator
that runs as a pre-processing step: invariants are automatically introduced in
the program and, although the implementation has some bugs, it strengths the
\textit{bkind} algorithm results.
Our results show that our \textit{bkind} with
invariants can considerably reduce the verification time of a large number of
benchmarks: in our experiments this is equivalent to almost two days reduction
in the verification time.


\begin{thebibliography}{10}

\bibitem{Biere:1999:SMC:646483.691738}
Biere, A., Cimatti, A., Clarke, E., Zhu, Y.:
\newblock {S}ymbolic {M}odel {C}hecking {W}ithout {BDD}s.
\newblock In: {T}ools {A}nd {A}lgorithms {F}or {T}he {C}onstruction {A}nd
  {A}nalysis {O}f {S}ystems. Volume 1633 of {LNCS}. (1999)  193--207

\bibitem{svcomp2017}
Beyer, D.:
\newblock {S}oftware {V}erification {W}ith {V}alidation {O}f {R}esults
  ({R}eport {O}n {SV}-{COMP} 2017).
\newblock In: {T}ools {A}nd {A}lgorithms {F}or {T}he {C}onstruction {A}nd
  {A}nalysis {O}f {S}ystems. Volume 10206 of {LNCS}. (2017)  331--349

\bibitem{Kroening2011}
Kroening, D., Ouaknine, J., Strichman, O., Wahl, T., Worrell, J.:
\newblock {L}inear {C}ompleteness {T}hresholds {F}or {B}ounded {M}odel
  {C}hecking.
\newblock In: {C}omputer-{A}ided {V}erification. Volume 6806 of {LNCS}. (2011)
  557--572

\bibitem{DBLP:books/daglib/0019162}
Bradley, A.R., Manna, Z.:
\newblock {T}he {C}alculus {O}f {C}omputation - {D}ecision {P}rocedures {W}ith
  {A}pplications {T}o {V}erification.
\newblock Springer (2007)

\bibitem{Beyer15}
Beyer, D., Dangl, M., Wendler, P.:
\newblock {B}oosting $k$-{I}nduction {W}ith {C}ontinuously-{R}efined
  {I}nvariants.
\newblock In: {C}omputer-{A}ided {V}erification. Volume 9206 of {LNCS}. (2015)
  622--640

\bibitem{Kinductor}
Donaldson, A., Haller, L., Kroening, D., R{\"{u}}mmer, P.:
\newblock {S}oftware {V}erification {U}sing $k$-{I}nduction.
\newblock In: {S}tatic {A}nalysis {S}ymposium. (2011)  351--368

\bibitem{Donaldson10}
Donaldson, A., Kroening, D., R{\"{u}}mmer, P.:
\newblock {SCRATCH}: {A} {T}ool {F}or {A}utomatic {A}nalysis {O}f {DMA}
  {R}aces.
\newblock In: {S}ymposium {O}n {P}rinciples {A}nd {P}ractice {O}f {P}arallel
  {P}rogramming. (2011)  311--312

\bibitem{EenS03}
E{\'{e}}n, N., S{\"{o}}rensson, N.:
\newblock {T}emporal {I}nduction {B}y {I}ncremental {SAT} {S}olving.
\newblock Electronic Notes in Theoretical Computer Science \textbf{89}(4)
  (2003)  543--560

\bibitem{GrosseLD09}
Gro{\ss}e, D., Le, H., Drechsler, R.:
\newblock {I}nduction-{B}ased {F}ormal {V}erification {O}f {S}ystem{C} {TLM}
  {D}esigns.
\newblock In: {W}orkshop {O}n {M}icroprocessor {T}est {A}nd {V}erification.
  (2009)  101--106

\bibitem{Sheera00}
Sheeran, M., Singh, S., St{\aa}lmarck, G.:
\newblock {C}hecking {S}afety {P}roperties {U}sing {I}nduction {A}nd {A}
  {SAT}-{S}olver.
\newblock In: {F}ormal {M}ethods {I}n {C}omputer-{A}ided {D}esign. (2000)
  108--125

\bibitem{GadelhaCN17}
Gadelha, M.Y.R., Monteiro, F.R., Cordeiro, L.C., Nicole, D.A.:
\newblock {T}owards {C}ounterexample-guided $k$-{I}nduction {F}or {F}ast {B}ug
  {D}etection.
\newblock In: {ACM} {J}oint {E}uropean {S}oftware {E}ngineering {C}onference
  {A}nd {S}ymposium {O}n {T}he {F}oundations {O}f {S}oftware {E}ngineering.
  (2018)

\bibitem{BISCHOFF200533}
Bischoff, G.P., Brace, K.S., Cabodi, G., Nocco, S.and~Quer, S.:
\newblock {E}xploiting {T}arget {E}nlargement {A}nd {D}ynamic {A}bstraction
  {W}ithin {M}ixed {BDD} {A}nd {SAT} {I}nvariant {C}hecking.
\newblock Electronic Notes in Theoretical Computer Science \textbf{119}(2)
  (2005)  33--49

\bibitem{Bradley13}
Hassan, Z., Bradley, A.R., Somenzi, F.:
\newblock {B}etter {G}eneralization {I}n {IC}3.
\newblock In: {F}ormal {M}ethods {I}n {C}omputer-{A}ided {D}esign, IEEE (2013)
  157--164

\bibitem{Jovanovic:2016:PK:3077629.3077648}
Jovanovi\'{c}, D., Dutertre, B.:
\newblock {P}roperty-directed $k$-induction.
\newblock In: {F}ormal {M}ethods {I}n {C}omputer-{A}ided {D}esign. (2016)
  85--92

\bibitem{DBLP:reference/mc/McMillan18}
McMillan, K.L.:
\newblock {I}nterpolation {A}nd {M}odel {C}hecking.
\newblock In: {H}andbook {O}f {M}odel {C}hecking.
\newblock Springer (2018)  421--446

\bibitem{kind-principle}
Wahl, T.:
\newblock {T}he $k$-induction {P}rinciple.
\newblock \url{http://www.ccs.neu.edu/home/wahl/Publications/k-induction.pdf}
  (2013) [Online; accessed September-2018].

\bibitem{MorseCNF13}
Morse, J., Cordeiro, L.C., Nicole, D.A., Fischer, B.:
\newblock Handling unbounded loops with {ESBMC} 1.20 - (competition
  contribution).
\newblock In: {TACAS}

\bibitem{Gadelha2015}
Gadelha, M.Y.R., Ismail, H.I., Cordeiro, L.C.:
\newblock {H}andling {L}oops {I}n {B}ounded {M}odel {C}hecking {O}f {C}
  {P}rograms {V}ia $k$-induction.
\newblock {International Journal on Software Tools for Technology Transfer}
  \textbf{19}(1) (2017)  97--114

\bibitem{Russell:2003:AIM:773294}
Russell, S.J., Norvig, P.:
\newblock {A}rtificial {I}ntelligence: {A} {M}odern {A}pproach. 2nd edn.
\newblock Pearson Education (2003)

\bibitem{10.1007/978-3-662-43376-8_3}
Cano, J., Delaval, G., Rutten, E.:
\newblock {C}oordination {O}f {ECA} {R}ules {B}y {V}erification {A}nd
  {C}ontrol.
\newblock In: {C}oordination {M}odels {A}nd {L}anguages, Berlin, Heidelberg,
  Springer Berlin Heidelberg (2014)  33--48

\bibitem{DBLP:conf/sbesc/RochaICB15}
Rocha, H., Ismail, H., Cordeiro, L.C., Barreto, R.S.:
\newblock {M}odel {C}hecking {E}mbedded {C} {S}oftware {U}sing $k$-{I}nduction
  {A}nd {I}nvariants.
\newblock In: {SBESC}. (2015)  90--95

\bibitem{DBLP:conf/tacas/RochaRIC017}
Rocha, W., Rocha, H., Ismail, H., Cordeiro, L.C., Fischer, B.:
\newblock {D}epth{K}: {A} $k$-{I}nduction {V}erifier {B}ased {O}n {I}nvariant
  {I}nference {F}or {C} {P}rograms - ({C}ompetition {C}ontribution).
\newblock In: {T}ools {A}nd {A}lgorithms {F}or {T}he {C}onstruction {A}nd
  {A}nalysis {O}f {S}ystems. (2017)  360--364

\bibitem{10.1007/978-3-319-89963-3_24}
Mal{\'i}k, V., Marti{\v{c}}ek, {\v{S}}., Schrammel, P., Srivas, M., Vojnar, T.,
  Wahlang, J.:
\newblock 2{LS}: {M}emory {S}afety {A}nd {N}on-termination.
\newblock In: {T}ools {A}nd {A}lgorithms {F}or {T}he {C}onstruction {A}nd
  {A}nalysis {O}f {S}ystems, Cham, Springer International Publishing (2018)
  417--421

\bibitem{Brain2015}
Brain, M., Joshi, S., Kroening, D., Schrammel, P.:
\newblock {S}afety {V}erification {A}nd {R}efutation {B}y $k$-{I}nvariants
  {A}nd $k$-{I}nduction.
\newblock In: {S}tatic {A}nalysis. (2015)  145--161

\bibitem{DBLP:conf/aaai/SturtevantF18}
Sturtevant, N.R., Felner, A.:
\newblock {A} {B}rief {H}istory {A}nd {R}ecent {A}chievements {I}n
  {B}idirectional {S}earch.
\newblock In: {C}onference {O}n {A}rtificial {I}nteligence, {AAAI} Press (2018)

\bibitem{cprover-manual}
Kroening, D.:
\newblock {CP}rover {M}anual.
\newblock \url{http://www.cprover.org/cprover-manual/} (2018) [Online; accessed
  September-2018].

\bibitem{CordeiroFM12}
Cordeiro, L.C., Fischer, B., Marques{-}Silva, J.:
\newblock {SMT}-{B}ased {B}ounded {M}odel {C}hecking {F}or {E}mbedded
  {ANSI}-{C} {S}oftware.
\newblock {IEEE} Transactions on Software Engineering \textbf{38}(4) (2012)
  957--974

\bibitem{esbmc2018}
Gadelha, M.R., Monteiro, F.R., Morse, J., Cordeiro, L.C., Fischer, B., Nicole,
  D.A.:
\newblock {ESBMC} 5.0: {A}n {I}ndustrial-{S}trength {C} {M}odel {C}hecker.
\newblock In: {A}utomated {S}oftware {E}ngineering, ACM (2018)  888--891

\bibitem{iridisSoton}
of~Southampton, U.:
\newblock {T}he {I}ridis {C}ompute {C}luster.
\newblock \url{https://www.southampton.ac.uk/isolutions/staff/iridis.page}
  (2018) [Online; accessed September-2018].

\bibitem{Clarke04}
Clarke, E., Kroening, D., Lerda, F.:
\newblock {A} {T}ool {F}or {C}hecking {ANSI}-{C} {P}rograms.
\newblock In: {T}ools {A}nd {A}lgorithms {F}or {T}he {C}onstruction {A}nd
  {A}nalysis {O}f {S}ystems. Volume 2988 of {LNCS}. (2004)  168--176

\end{thebibliography}

\end{document}